\documentclass[aps,preprint,amsmath,amssymb,showpacs,superscriptaddress,nofootinbib]{revtex4-1}

\usepackage{graphicx,amssymb,amsmath,float}
\usepackage{gensymb}
\usepackage{hyperref}
\usepackage{slashed}
\usepackage[utf8]{inputenc}

\usepackage{titlesec}

\hypersetup{
    colorlinks=false,
    linktocpage,
    }

\begin{document}
\title{The SuSA model for neutrino oscillation experiments: from quasielastic  scattering to the resonance region}
 \author{M.B.~Barbaro} \affiliation{Dipartimento di Fisica,
   Universit\`{a} di Torino, Via P. Giuria
   1, 10125 Torino, Italy} 
   \affiliation{Istituto Nazionale di Fisica Nucleare, Sezione di Torino, Via P. Giuria
   1, 10125 Torino, Italy}
   \affiliation{Universit\'e Paris-Saclay, CNRS/IN2P3, IJCLab, 91405 Orsay, France}
\author{A.~De Pace} \affiliation{Istituto Nazionale di Fisica Nucleare, Sezione di Torino, Via P. Giuria
   1, 10125 Torino, Italy}
\author{L.~Fiume} \affiliation{Dipartimento di Fisica,
   Universit\`{a} di Torino, Via P. Giuria
   1, 10125 Torino, Italy}

\date{\today}

\begin{abstract}
High precision studies of Beyond-Standard-Model physics through accelerator-based neutrino oscillation experiments require a very accurate description of neutrino-nucleus cross sections in a broad energy region, going from quasielastic scattering up to deep inelastic scattering. In this work we focus on the following processes: quasielastic scattering, two-particle-two-hole excitations, 
and the excitation of the first (Delta) and second (Roper) resonances of the nucleon. The nuclear model is fully relativistic and includes both one- and two-body currents. We compare our results with recent T2K and MicroBooNE data on carbon and argon targets, and present predictions for DUNE kinematics. 
\end{abstract}


\maketitle


\label{sec:intro}
\section{Introduction}


The accurate description of neutrino-nucleus cross sections in the GeV regime is essential for the interpretation of present and future neutrino oscillation experiments, aimed at  precision measurements of the neutrino properties and at the search of physics beyond the Standard Model~\cite{Alvarez-Ruso:2017oui}. In particular, the future HyperK~\cite{Abe:2015zbg} and DUNE~\cite{Abi:2020wmh} facilities are expected to measure the leptonic CP-violating phase $\delta_{CP}$, which could shed light on the origin of  the matter/antimatter asymmetry in the Universe. Encouraging results in this direction have recently been  published by the T2K collaboration~\cite{Abe:2019vii}.

     The extraction of the oscillation parameters entering the neutrino mixing matrix $U$ from the measurements of the oscillation probabilities between  different flavours crucially depends on the precise knowledge of the neutrino energy, which must be inferreded from the kinematics of the detected particles in the final state.  Detectors are made of heavy nuclei (carbon, oxygen, argon) and  a large part of the systematic error in the experimental analyses comes  from modelling of neutrino-nucleus interaction. The success of future experiments relies  on the ability of reducing these nuclear uncertainties  in a wide energy range, from the quasielastic (QE) region, corresponding to the elastic interaction of the neutrino with a single nucleon inside the nuclear target, up to the deep inelastic scattering (DIS) domain, where the probe interacts with the constituent quarks.
     
While the QE region  has been extensively studied in recent years by various groups~\cite{Amaro:2006tf,Martini:2011wp,Martini:2009uj,Martini:2010ex,Nieves:2011yp,Meucci:2014bva,Meucci:2011vd,Pandey:2014tza,Ivanov:2018nlm,Ivanov:2013saa,Rocco:2016ejr,RuizSimo:2018kdl,Amaro:2011qb,Benhar:2010nx,Rocco:2018mwt}, the resonance region between the QE and the DIS regimes still needs to be fully investigated~\cite{SajjadAthar:2020nvy,Alvarez-Ruso:2017oui}. This region corresponds to the excitation of nucleon resonances and will play a major role in the  kinematic domain explored by  DUNE. Moreover, an important contribution to the cross section arises from the excitation of two-particle-two-hole (2p2h) states, which occurs at kinematics between the QE and the $\Delta$-resonance peaks and is induced by meson-exchange currents~\cite{Martini:2012fa,Gran:2013kda,Simo:2016ikv,Amaro:2011aa,Amaro:2010sd}.

Nuclear models to be used in this context must satisfy some basic requirements. First of all,  since typical energies belong to the GeV region, they must be relativistic or at least contain relativistic corrections. The simplest fully relativistic nuclear model is the global Relativistic Fermi Gas (RFG), which constitutes a solid basis for more sophisticated models. The RFG  framework allows for an exact relativistic treatment of both currents and nuclear states, but ignores NN correlations, apart from the statistical ones embodied in the Pauli principle. A semi-phenomenological improvement of the RFG model is represented by the SuSA (Super Scaling Approximation) model, which takes into account both initial and final state interactions as extracted from the analysis of electron scattering data at different kinematics and on different nuclei~\cite{Amaro:2004bs,Amaro:2019zos}. Another fully relativistic model, the relativistic mean field (RMF), has been shown to explain from the microscopic point of view the basic features of the SuSA approach and has been used to build an updated version of the model (SuSAv2)~\cite{Gonzalez-Jimenez:2014eqa}, which has been applied to the study of neutrino reactions in the quasielastic region~\cite{Megias:2014qva,Megias:2016fjk}.

A second important feature required from a reliable nuclear model is consistency: the different kinematic regions and elementary processes should be described within the same theoretical framework. Consistency is easily accomplished in the RFG model, but difficult to achieve in more sophisticated models. For example, the available calculations of the 2p2h response are mostly performed in the RFG framework, and combining them with other contributions evaluated using different, although more sophisticated, nuclear models, may lead to misleading or incorrect results.

In this paper we will focus in particular on the QE, 2p2h and resonance regions - the latter including the first (Delta) and second (Roper) excited states of the nucleon - within the RFG and SuSA models. The results will be compared with recent neutrino data from the T2K and MicroBooNE experiments and predictions will be shown for typical DuNE kinematics. 

The paper is organized as follows: in Section \ref{sec:ccnni} we summarize the formalism for charged current neutrino nucleus reactions induced by one- and two-body currents. The nuclear model described in Section \ref{sec:models} is used to derive the results presented in Section \ref{sec:results}, where we compare the theoretical predictions with experimental data. Finally, in Section \ref{sec:concl} we draw our conclusions and outline the future deveopments of this research.

\section{Charged current neutrino-nucleus interactions}
\label{sec:ccnni}

Let us consider the $(\nu_l,l^-)$ charged-current (CC) cross section for the process
\begin{equation}
\nu_l+A \longrightarrow l^- + X \,,
\end{equation}
where a neutrino with given energy $E_\nu$ and momentum $\vec k$ hits a nucleus $A$ and a negative charge lepton $l^-$ is detected in the final state with energy $E_l$, momentum $\vec k'$ and scattering angle $\theta_l$. Here $X$ can be any unobserved hadronic system, containing one or more knocked out nucleons, pions and other mesons, etc.  
The corresponding cross section is obtained from the contraction of the  leptonic and hadronic tensors. The latter encodes the full dependence on the nuclear dynamics and is defined in the target rest frame as 
\begin{equation}
W^{\mu\nu}(\vec q,\omega) = \sum_n <A|J^{\mu\dagger}|n><n|J^\nu|A> \,\delta(\omega+E_A-E_n)\,\delta(\vec q-\vec p_n)\,,
\label{eq:Wmunu}
\end{equation}
where $J^\mu$ is the weak hadronic current, $|A>$ is the initial nuclear ground state having energy $E_A$ and $|n>$ are all the  intermediate nuclear states, of energy $E_n$ and momentum $\vec p_n$, accessible through the current operator.  The $\delta$ functions express energy and momentum conservation, $\omega=E_\nu-E_l$ and $\vec q=\vec k-\vec k'$ being the energy and momentum transferred from the probe to the hadronic system.
The double differential cross section  can be  expressed 
as the linear combination of five response functions~\cite{Amaro:2004bs}
\begin{equation}
\frac{d^2\sigma}{dE_l d\cos\theta_l}
=
\sigma_0
\left(
V_{CC} R_{CC}+
2{V}_{CL} R_{CL}
+{V}_{LL} R_{LL}+
{V}_{T} R_{T}
+
2{V}_{T'} R_{T'}
\right) \, , 
\end{equation}
where 
\begin{equation}
\sigma_0= 
\frac{G^2\cos^2\theta_c}{4\pi}
\frac{k'}{E_l} \left[(E_\nu+E_l)^2-\vec q^2\right]\,, 
\label{eq:sig0}
\end{equation}
being $G=1.166\times
10^{-11}$ MeV$^{-2} $  the Fermi weak constant and
$\cos\theta_c=0.975$ the Cabibbo angle.
The coefficients $V_K$ depend only on the lepton kinematics and are defined in Ref.~\cite{Amaro:2004bs}, while the response functions $R_K\equiv R_K(|\vec q|,\omega)$, also defined in Ref.~\cite{Amaro:2004bs}, depend only on the three-momentum $\vec q$ and energy $\omega$ transferred to the nucleus. The indices $C$, $L$, $T$ refer to the Coulomb, longitudinal and transverse components of the leptonic and hadronic currents
with respect to  $\vec q$. The response functions 
\begin{eqnarray}
&&R_{CC } =  W^{00 } \,,\ 
R_{CL }  =  -\frac12 (W^{03 } + W^{30 } ) \,,\ 
R_{LL }  =  W^{33 }\,,  \\
&&R_{T } =  W^{11 } + W^{22 } \,,\ 
R_{T' }  =  -\frac{i}{2}(W^{12 } - W^{21 })
\end{eqnarray}
are specific components of the hadronic tensor \eqref{eq:Wmunu}, which includes both one- and two-body terms.

\subsection{One-body hadronic tensor}
\label{sec:elres}

The elastic $N\to N$ and resonance production $N\to N^*$  processes are induced by one-body currents and can be treated simultaneously by introducing  the inelasticity parameter~\cite{Amaro:2004bs},\begin{equation}
\rho=1-\frac{m^{*2}-m^2}{q^2}\,,
\end{equation}
where $m$ and $m^*$  are the nucleon and resonance mass, respectively, and $q^2=\omega^2-\vec q^2$ the squared four-momentum transfer. 
In the elastic case $m^*=m$ and $\rho=1$.
The single-nucleon tensor  can be written in the general form~\cite{Donnelly:2017aaa}
\begin{eqnarray}
w_{1b}^{\mu\nu}&=& -w_{1} \left(g^{\mu\nu}-\frac{q^\mu q^\nu}{q^2}\right) 
+ \frac{w_{2}}{m^2} \left(p^\mu+\frac{\rho}{2}\,q^\mu\right) \left(p^\nu+\frac{\rho}{2}\,q^\nu\right)
+i \frac{w_3}{m^2}\epsilon^{\alpha\beta\mu\nu}p_\alpha q_\beta
\nonumber\\
&+& \frac{u_1}{q^2} q^\mu q^\nu+\frac{u_2}{2 m^2}(p^\mu q^\nu+q^\mu p^\nu)\,,
\label{eq:wmunu-sn}
\end{eqnarray}
 where the structure functions $w_i$ and $u_i$ depend on the specific process and are evaluated starting from the transition current.

In this work we take into account  the first two excited states of the nucleon, which dominate at the kinematics we are exploring: the spin 3/2, isospin 3/2, $P_{33}(1232)$ ($\Delta$) resonance and the spin 1/2 , isospin 1/2, $P_{11}(1440)$ (Roper) resonance. Higher resonances can be easily included in the calculation.

The structure functions relative to elastic scattering and to the  $N\to\Delta$ transition are given in Ref.~\cite{Amaro:2004bs} and will not be repeated here. 

In the Roper resonance region the  
$N\to P_{11}(1440)$  weak current is~\cite{Lalakulich:2006sw}
 \begin{equation}
J^\mu(N\to P_{11}) = \Gamma^\mu_V-\Gamma^\mu_A \,,
\end{equation}
 where 
  \begin{eqnarray}
 \Gamma^\mu_V &=& 2F_1^*\left(\gamma^\mu - \frac{{\not \! q} q^\mu}{q^2} \right) + \frac{2F_2^*}{m+m^*} i\sigma^{\mu\nu}q_\nu \,,\\
 \Gamma^\mu_A &=& G_A^*\gamma^\mu\gamma_5 + \frac{G_P^*}{2m} q^\mu\gamma_5 
 \end{eqnarray}
are  the vector and axial operators.
 The $N\to P_{11}$ transition form factors $F_i^*$ and $G_i^*$ are given in Appendix A.
 
 The corresponding single nucleon tensor is given by
 \begin{equation}
 w^{\mu\nu}_{N\to P_{11}} = \frac{1}{2} \frac{m^*}{m} Tr\left\{\frac{{\not\! p}+m}{2m} \left(\Gamma^\mu_V-\gamma_0\Gamma^{\mu\dagger}_A\gamma_0\right)\frac{{\not\! p'}+m^*}{2m^*}\left(\Gamma^\nu_V-\Gamma^\nu_A\right)\right\}
 \end{equation}
 and, after a lengthy calculation, can be recast in the form \eqref{eq:wmunu-sn} with the following the $N\to P_{11}$ structure functions
\begin{eqnarray}
w_{1} &=&4 \left(F_1^*+F_2^*\right)^2 \left[\tau+\left(\frac{\mu^*-1}{2}\right)^2\right] + G_A^{*2} (1+\tau^*) \left(\frac{\mu^*+1}{2}\right)^2 \,,\\
w_{2} &=& (2 F_1^*)^2 +\tau^* (2 F_2^*)^2 + G_A^{*2} \,,\\
w_3 &=& 2 (F_1^*+F_2^*) G_A^* \,,\\
u_1 &=& -\tau G_A^{*2} \left(\frac{1}{\tau^*}+1-\rho^2\right)+2\tau G_A^* G_P^*-\tau G_P^{*2} \left[\tau+\left(\frac{\mu^*-1}{2}\right)^2\right] \,,\\
u_2&=&G_A^{*2} (1-\rho) +  G_A^* G_P^* \left(\frac{\mu^*-1}{2}\right) \,,
\end{eqnarray}
where $\mu^*\equiv m^*/m$, $\tau=-q^2/(4m^2)$ and $\tau^*=-q^2/(m+m^*)^2$.

\subsection{Two-body hadronic tensor}
\label{sec:2p2h}

Processes induced by two-body currents correspond to the interaction of the neutrino with a pair of correlated nucleons, leading to a 2p2h final state in which two nucleons are knocked out of the nuclear ground state. The nucleon-nucleon correlations can be modelled through the exchange of a meson and the resulting meson-exchange currents (MEC) are largely dominated by the pion. The kinematical region in which such processes occur corresponds to energy transfers between the quasielastic and $\Delta$ resonance peaks, where the MEC are known to be essential in order to describe inclusive electron scattering data~\cite{DePace:2003spn,Donnelly:1978xa,Megias:2016lke}.

\begin{figure}[H]
\center
\includegraphics[width=10.5 cm]{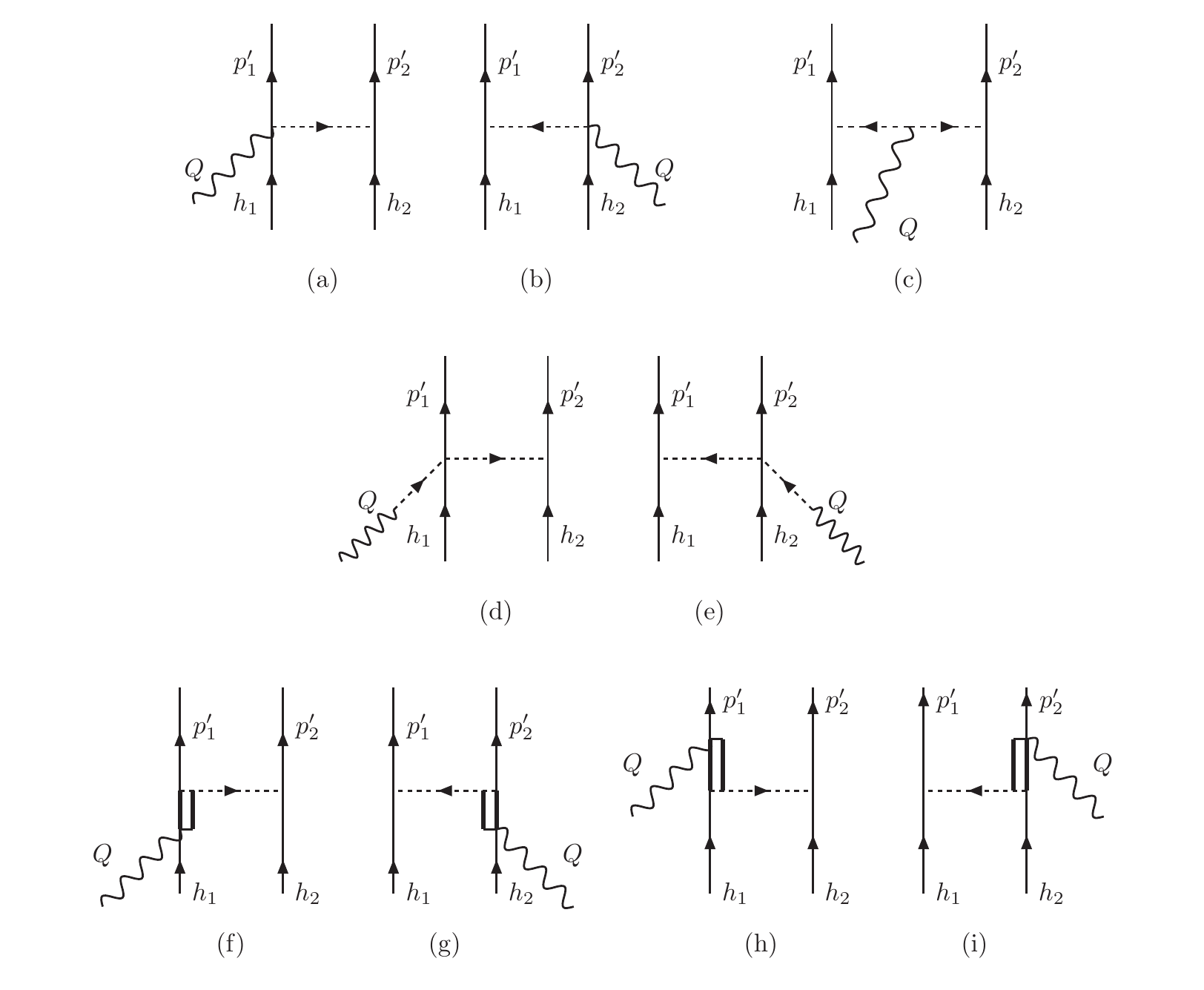}
\caption{Weak meson exchange currents considered in this work.\label{fig:MEC}}
\end{figure}

The diagrams contributing to the weak pionic MEC in the vacuum are shown in Fig.~\ref{fig:MEC} and are usually classified as contact (a,b), pion-in-flight (c), pion pole (d-e) and $\Delta$-MEC (f-i). Explicit expressions for the two-body tensor $w_{2b}^{\mu\nu}$ for neutrino scattering can be found in Ref.~\cite{Simo:2016ikv}.

As a next step one needs to embed the above one- and two-body elementary tensors into a nuclear model.

\section{The SuSA model}
\label{sec:models}

The simplest approach to a fully relativistic nuclear system is represented by the Relativistic Fermi Gas (RFG) model, in which the single-nucleon wave functions are free plane waves multiplied by Dirac spinors and the only correlations are the statistical ones induced by the Pauli principle. Each nucleus is characterized by a Fermi momentum $k_F$, usually fitted to the width of the quasielastic peak in electron scattering data.

The  one-body nuclear tensor, in both the quasielastic and the resonance regions, is given by
\begin{equation}
W^{\mu\nu}_{1b,RFG}(\vec q,\omega;\rho) 
= \int d\vec p\, \delta(E'-E-\omega)
\frac{m^2}{EE'} 
\,w^{\mu\nu}_{1b}(\vec p+\vec q,\vec p;\rho)
\,\theta(k_F-p) \,,
\label{eq:wmunu}
\end{equation}
where $(E,\vec p)$ and $(E',\vec p+\vec q)$ are the on-shell energies and momenta of the initial and final hadrons, respectively, and $w^{\mu\nu}_{1b}$ is the  elementary single-nucleon tensor defined in Eq.~\eqref{eq:wmunu-sn}. 
In the quasielastic case $\rho$=1 and an extra $\theta(|\vec p+\vec q|-k_F)$  must be inserted inside the integral \eqref{eq:wmunu} to account for the Pauli exclusion principle.

In this model the response functions can be evaluated analytically and can be expressed in the general form
\begin{equation}
R^K(|\vec q|,\omega) = U^K(|\vec q|,\omega) f_{RFG}(\psi_\rho(|\vec q|,\omega;k_F))\,,
\end{equation}
where the $U^K$ are functions that depend  on the nucleon-boson vertex and 
incorporate corrections due to the Fermi motion, while the "superscaling" function
\begin{equation}
f_{RFG}(\psi)=\frac{3}{4}(1-\psi^2)\,\theta(1-\psi^2)
\label{eq:fRFG}
\end{equation}
is a universal function - namely valid for all the one-body responses - depending only on one scaling variable $\psi(|\vec q|,\omega; k_F)$. The latter is a specific combination of the transferred energy and momentum given by
\begin{equation}
\psi_\rho(|\vec q|,\omega; k_F) = \frac{1}{\sqrt{\xi_F}} \frac{\lambda-\tau\rho}{\sqrt{(1+\lambda\rho)\tau+\kappa\sqrt{\tau(1+\tau\rho^2)}}} \,,
\end{equation}
with $\xi_F=\sqrt{(k_F/m)^2+1}-1$; $\lambda=\omega/(2m)$ and $\kappa=|\vec q|/(2m)$ are dimensionless Fermi kinetic energy, energy transfer and momentum transfer, respectively. Physically the scaling variable $\psi_\rho$ represents, in the model, the minimal kinetic energy of the initial state nucleons  participating to the reaction at given $|\vec q|$ and $\omega$ in a nucleus characterized by the Fermi momentum $k_F$.

The RFG model has the advantage of being relativistic and therefore represents a suitable starting point for more sophisticated models, but it is well-known that it gives a poor description of electron scattering data. These, unlike neutrino data, are very abundant and precise and can be used as a benchmark in neutrino scattering studies. It was first suggested in Ref.~\cite{Amaro:2004bs} that the scaling behaviour of $(e,e')$ data can also be used as an input to get reliable predictions for neutrino-nucleus cross sections. This idea is at the basis of the SuSA model, which essentially amounts to replace the RFG superscaling function \eqref{eq:fRFG} by a phenomenological one, $f_{SuSA}(\psi)$, extracted by the analysis of electron scattering data as the ratio between the double differential cross section and an appropriate single-nucleon function~\cite{Donnelly:1998xg,Donnelly:1999sw}. 
The analysis of the longitudinal quasielastic data shows that this function is very weakly dependent on the momentum transfer $\vec q$ providing the latter is high enough (namely larger than about 400 MeV/c) to allow for the impulse approximation; this property is usually referred to as scaling of first kind. Moreover, the superscaling function is almost independent of the specific nucleus for mass numbers $A$ ranging from 4 (helium) up to 198 (gold); this is known as scaling of second kind. Superscaling is the simultaneous occurrence of the two kinds of scaling and is well respected by electron scattering data in the QEP region. Scaling violations occur in the transverse channel due to non-impulsive contributions like 2p2h excitations. 

The phenomenological superscaling function $f_{SuSA}$ incorporates effectively NN correlations and final state interactions and gives, by construction, a good agreement with $(e,e')$ data in a wide range of kinematics and mass numbers. 
The parametrization used in this work is
\begin{equation}
f_{SuSA}(\psi) = \frac{\alpha}{\left[1+\beta^2\left(\psi+\gamma\right)^2\right] \left(1+e^{-\delta\psi}\right)} \,,
\label{eq:fSuSA}
\end{equation}
where the parameters are fitted to the electron scattering quasielastic world data analyzed in Ref.~\cite{Jourdan:1996np,Donnelly:1999sw} for all the experimentally available kinematics and nuclear targets.
Here we use the values $\alpha=2.9883$, $\beta=1.9438$, $\gamma=0.6731$ and $\delta=3.8538$, corresponding to the fit performed in Ref.~\cite{Maieron:2001it}.
Two more parameters, the Fermi momentum  $k_F$ (228 MeV/c for carbon and 241 MeV/c for argon) and the energy shift $E_s$ (20 MeV), are fitted for each nucleus to the experimental width and position of the quasielastic peak~\cite{Maieron:2001it}. 

\begin{figure}[H]
\center
\includegraphics[width=7.5 cm]{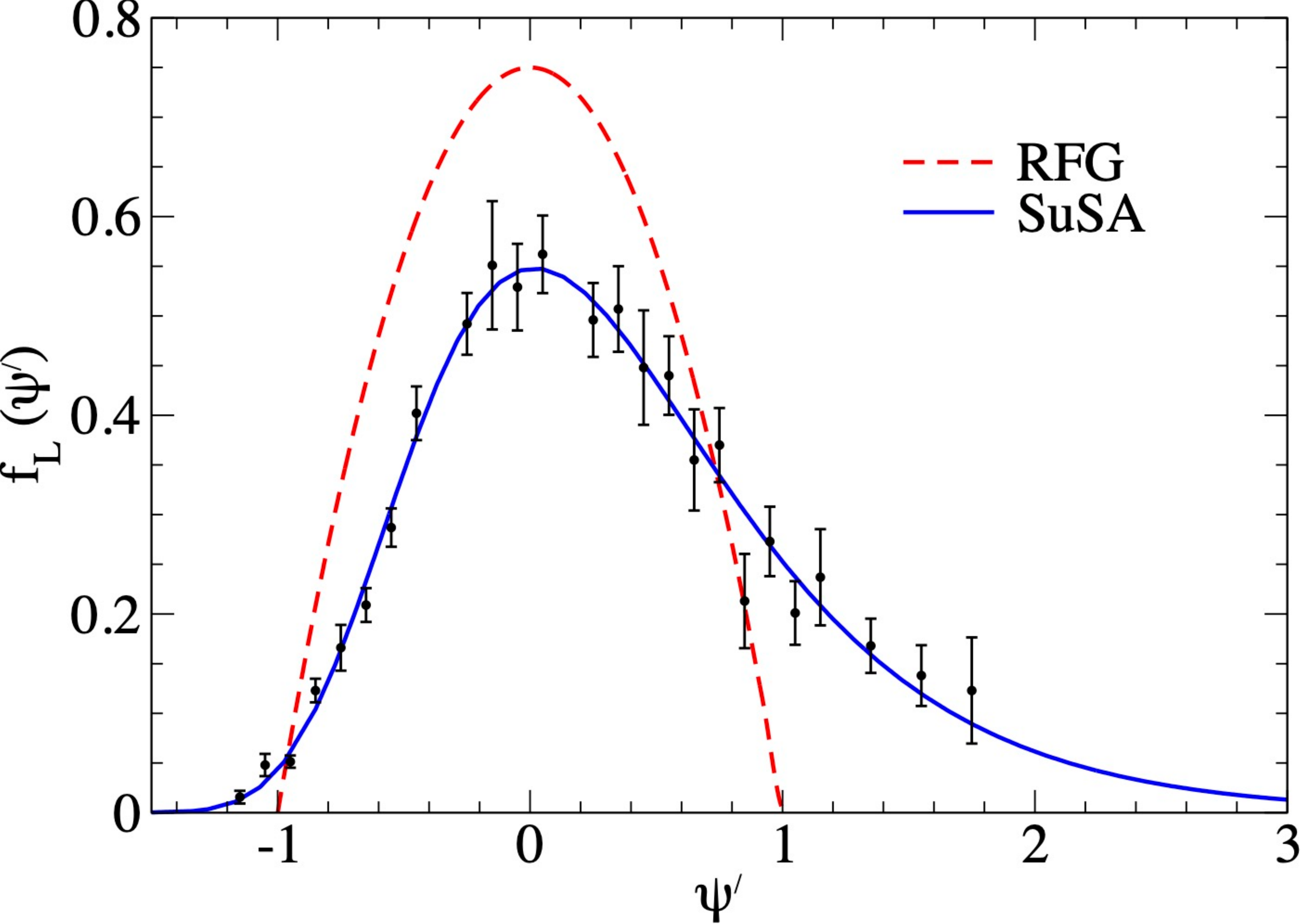}
\caption{The RFG and SuSA scaling functions compared with the world averaged longitudinal inclusive electron scattering data~\cite{Jourdan:1996np}. \label{fig:scaling}}
\end{figure}

In Fig.~\ref{fig:scaling} the RFG and SuSA scaling functions, Eqs.~\eqref{eq:fRFG} and \eqref{eq:fSuSA}, are compared with the world averaged longitudinal (e,e') data~\footnote{Here the scaling variable is defined as $\psi'=\psi((|\vec q|,\omega-E_s; k_F)$ to incorporate the energy shift $E_s$.}. This comparison clearly shows that the RFG provides a rather poor description of electron scattering data and more realistic models must be applied to neutrino oscillation analyses.
Note that, although the scaling function $f_{SuSA}$ has been extracted from quasielastic data, in the present work we assume it to be valid also in the resonance production region.  This choice is motivated on the one hand by the RFG result, for which the universality of the superscaling function is exactly true, and on  the other by the fact that the nuclear effects embodied in $f_{SuSA}$ are expected to depend not too strongly on the reaction channel.
 The superscaling function embodies nuclear effects which account for both initial and final state physics. It is reasonable to assume that the initial state physics, essentially described by the nuclear spectral function, is independent of the reaction channel. On the other hand, the final state interactions of the produced hadrons with the nuclear medium in principle distort the scaling function in a different way in each channel. However, it was shown in Refs.~\cite{Megias:2016lke,Barbaro:2019vsr} that the use of a universal scaling function in the full spectrum provides a good description of electron scattering data in a wide kinematical range. This makes us confident that the error associated to this approximation is not too large when the model is applied to neutrino scattering.
 It is also worth mentioning that an alternative approach has been taken in Refs.~\cite{Amaro:2004bs,Maieron:2009an,Ivanov:2015aya}, where a scaling function to be used in the $\Delta$ resonance region, different from the quasielastic one, has been extracted from electron scattering data. This method provides a phenomenological description valid at transferred energies below the $\Delta$ peak, while at higher $\omega$ it fails due to the opening of other inelastic channels.

Studies on the microscopic origin  of  the scaling function have shown that the shape and size of $f_{SuSA}$ can be reproduced with good accuracy by the relativistic mean field model~\cite{Caballero:2005sj}. In particular, it was shown that the high-energy asymmetric tail displayed by $f_{SuSA}$ can be mainly ascribed to  final state interactions and it cannot be reproduced if the latter are neglected (plane wave impulse approximation) or treated inconsistently with the initial state (like for instance using an optical potential). 
The RMF model was also exploited to construct a new version of the superscaling model (SuSAv2)~\cite{Gonzalez-Jimenez:2014eqa,Megias:2016lke}, where different scaling functions are used in each channel (longitudinal, transverse and axial, isoscalar and isovector), as predicted by the model in the quasielastic region. Although the differences between SuSA and SuSAv2 are not negligible, in this paper we stick to the original SuSA model, which employs the same scaling function in all channels and treats consistently the quasielastic and inelastic processes. Further refinements of the model will be explored in future work.

The superscaling approach above described is based on the assumption that the neutrino interacts with a single nucleon (impulse approximation) and ignores  the interaction of the probe with two correlated nucleons. These processes violate scaling of both kinds~\cite{DePace:2004cr} and obey a different scaling law, theoretically predicted  in Ref.~\cite{Amaro:2017eah} and well respected by experimental (e,e') data from different nuclei~\cite{Dai:2018gch,Barbaro:2019vsr}. They are added to the model within the RFG framework.

The two-body nuclear tensor corresponding to the MEC previously introduced is evaluated in the RFG model as
 \begin{eqnarray}
W^{\mu\nu}_{2b, RFG}
&=&\frac{V}{(2\pi)^9}\int
d\vec p_1
d\vec p_2
d\vec h_1
d\vec h_2
\frac{m^4}{E_{h_1}E_{h_2}E_{p_1}E_{p_2}}
w^{\mu\nu}_{2b}(\vec p_1,\vec p_2,\vec h_1,\vec h_2)
\nonumber\\
&\times&
\theta(|\vec p_1|-k_F)\theta(|\vec p_2|-k_F)\theta(k_F-|\vec h_1|)\theta(k_F-|\vec h_2|)
\nonumber\\
&\times&
\delta(E_{p_1}+E_{p_2}-E_{h_1}-E_{h_2}-\omega)
\delta(\vec p_1+\vec p_2-\vec q-\vec h_1-\vec h_2) \, ,
\label{wmunu2b}
\end{eqnarray}
where an integral appears over all
the 2p2h excitations of the RFG with two holes ($\vec h_1$, $\vec h_2$) and
two particles ($\vec p_1$,$\vec p_2$)  in the final state, and
 $w^{\mu\nu}_{2b}$ is the elementary two-body tensor represented in Fig.~\ref{fig:MEC} (see Ref.~\cite{Simo:2016ikv}).

The computation of the 2p2h responses in the RFG is time consuming due to the high dimensionality (7) of the integrals. For the purpose of the present work, where an extra integral over the neutrino flux must be performed before comparing the results to the experimental data, we make use of a parametrization of the  numerical results obtained in Refs.~\cite{Megias:2016fjk,Megias:2014qva}. This parametrization gives a very accurate representation of the exact results in a wide kinematic range (momentum transfers up to 2 GeV/c) and provides an efficient way of getting completely equivalent results.

Further details on the SuSA+MEC model and on the connection between electron and neutrino scattering can be found in  the recent review article \cite{Amaro:2019zos}.

\section{Results}
\label{sec:results}

We now present the predictions of the model introduced in the previous Section and compare them to some recent experimental data.
We consider two kinds of CC $\nu_\mu$-nucleus data. The first are "$0\pi$" (or "QE-like") data, where only the outgoing muon is detected and the final state does not contain pions. These data are supposed to correspond mainly to quasielastic scattering (one nucleon knockout) and to 2p2h excitations (two nucleons knockout). Note that in the latter the pion exchanged between the correlated nucleons is always highly virtual. The second set of data is instead of inclusive type, in the sense that, again, only the final lepton is detected but the final state can contain any unobserved hadrons (one or more nucleons, pions, other mesons). In this case the cross section receives contribution not only from the QE and 2p2h processes, but also from the excitation of nucleon resonances, which subsequently decay into undetected nucleons and mesons. The non-resonant meson production can also contribute to the signal, but is supposed to be less important, as suggested by the results of Refs.~\cite{Hernandez:2007qq,Amaro:2008hd,Rein:1980wg,Lalakulich:2010ss}, and will therefore be ignored in this work.

\begin{figure}[H]
\center
\includegraphics[width=8.5 cm]{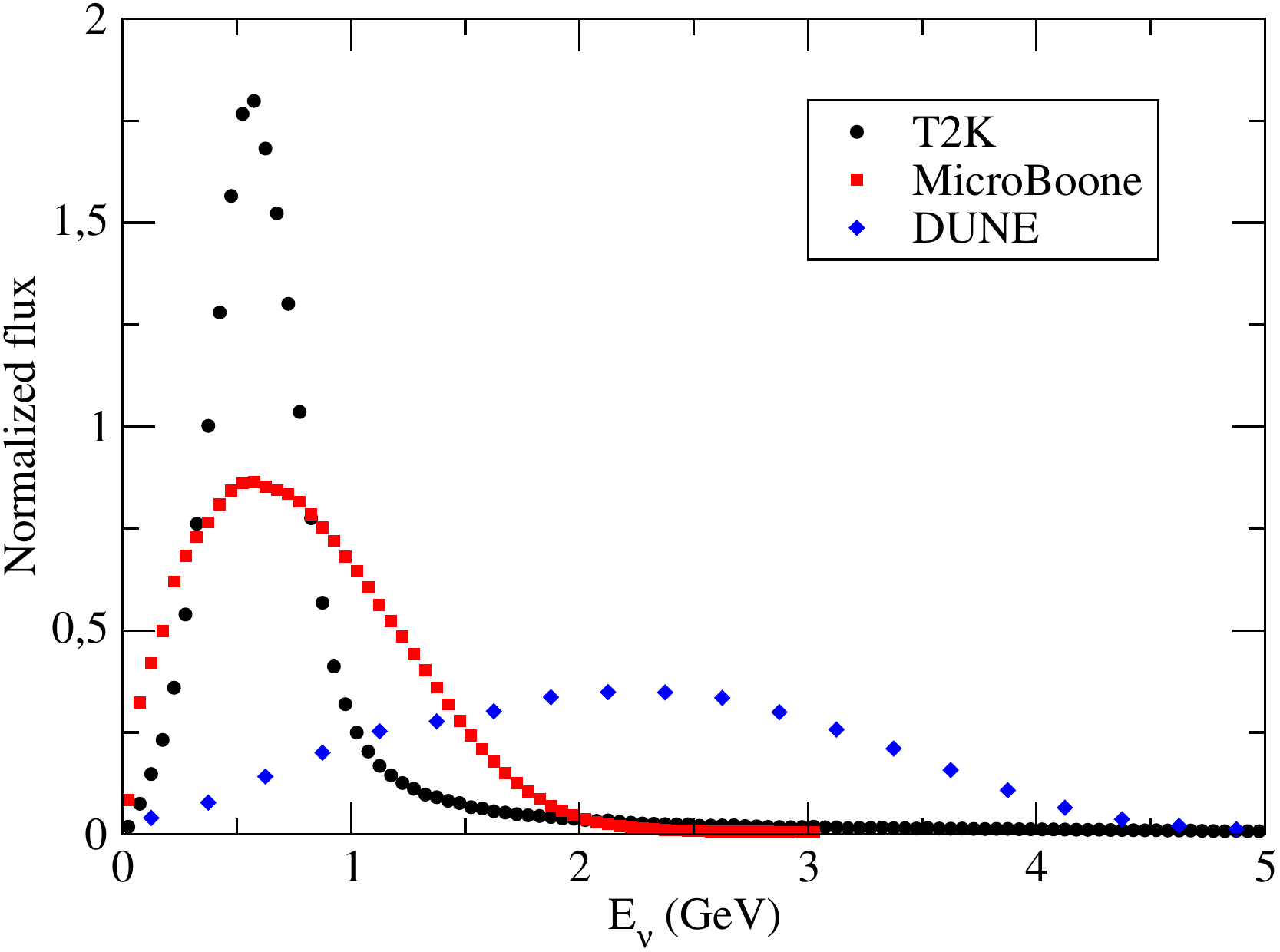}
\caption{The normalized T2K (off-axis)~\cite{PhysRevD.91.072010}, MicroBooNE~\cite{MicroBooNE:2018efi} and DUNE~\cite{DuneFlux} muon-neutrino fluxes displayed versus the neutrino energy $E_\nu$. \label{fig:fluxes}}
\end{figure}

We first compare our results with data published by the T2K~\cite{Abe:2016tmq,Abe:2013jth} and MicroBooNE~\cite{Abratenko:2019jqo} collaborations. Although the two experiments explore similar kinematics, the T2K off-axis neutrino flux is more focused than the broader MicroBooNE flux (see Fig.~\ref{fig:fluxes}) and this may have consequences on the relative contributions of different processes. Moreover, the nuclear targets are different: carbon for T2K and argon for MicroBooNE. This also can induce differences in nuclear effects that depend on the nuclear density.

The Fermi momenta employed in this work are $k_F$=228 MeV/c and 241 MeV/c for carbon and argon, respectively, and the energy shift $E_s$=20 MeV. These values were fitted to inclusive electron scattering data in Ref.~\cite{Maieron:2001it}.

Before showing the results, a comment is in order concerning Pauli blocking effects. As already mentioned, in the low $(\omega,|\vec q|)$ regime, where these effects come into play,  approaches based on the impulse approximation like the RFG and SuSA models should be taken with care. Nevertheless, since neutrino data also include this region, we include Pauli blocking in the SuSA model following the procedure originally proposed  in Ref.~\cite{Rosenfelder:1980nd}. This generalizes the RFG prescription $|\vec p+\vec q|>k_F$, valid only for a step-like momentum distribution, and amounts to the following replacement for the superscaling function
\begin{equation}
f(\psi(\omega)) \longrightarrow f(\psi(\omega)) - f(\psi(-\omega))\,.
\end{equation}

\begin{figure}[H]
\center
\includegraphics[height=12.5 cm, width=10.5 cm]{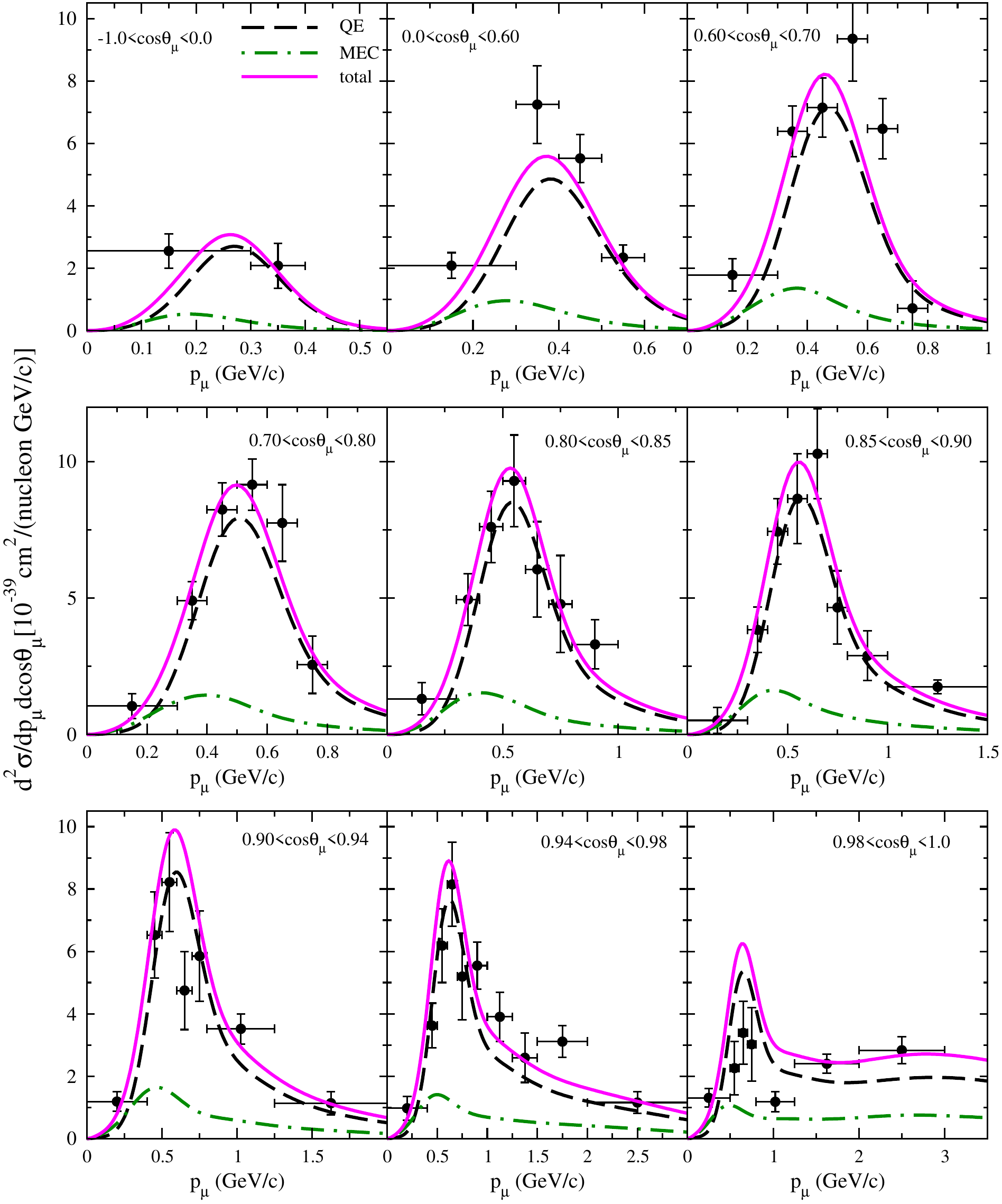}
\caption{The SuSA  double differential $(\nu_\mu,\mu^-)$ cross section off $^{12}C$ with no pions in the final state, averaged over the T2K flux, is displayed versus the muon momentum $p_\mu$. The separate QE and MEC contributions are also shown.
The data correspond to Analysis 1 from Ref.~\cite{Abe:2016tmq}. \label{fig:T2K0pi}}
\end{figure}

In Fig.~\ref{fig:T2K0pi} we show the SuSA model predictions for the T2K double differential $(\nu_\mu,\mu^-)$ cross section off $^{12}C$ with no pions in the final state as a function of the muon momentum $p_\mu$, for different bins of the scattering angle $\theta_\mu$. The separate QE and MEC contributions are also shown. In all cases the contribution of MEC (2p2h excitations) is sizable and necessary in order to explain the experimental data.  The agreement with the data is rather good, except for the last angular bin and low $p_\mu$, corresponding to very small values of scattering angle. This is not surprising since at these kinematic conditions, where small values of the energy  and momentum transfer  play a major role, 
superscaling ideas are not applicable and, in general, any model based on the impulse approximation is hardly reliable. In this region nuclear collective effects can take place and different approaches, like the one based on RPA, are more appropriate.

It should also be mentioned that the $CC0\pi$ cross section could also receive contribution  from pion production followed by re-absorption in the nucleus, a process not included in our calculation. This would require the microscopic description of pion production and its final state interactions, which is not available at present in our phenomenological model, where FSI are effectively absorbed into the scaling function. According to NEUT~\cite{Hayato:2009zz} and GENIE~\cite{ANDREOPOULOS201087} Monte Carlo generators this contribution accounts for about 10\% of the neutrino measured cross section~\cite{Abe:2020jbf}. It should be added to the theoretical calculation, or subtracted from the data, for a detailed quantitative comparison, which is beyond the scope of this work.

Having validated the QE and MEC model versus $0\pi$ data, we now compare our results with inclusive data, which get contribution also from inelastic channels. As previously stated, in our approach we include the excitation of the first two nucleon resonances, the $P_{33}(1232)$ ($\Delta$) and the $P_{11}(1440)$ (Roper).

\begin{figure}[H]
\center
\includegraphics[width=12.5cm]{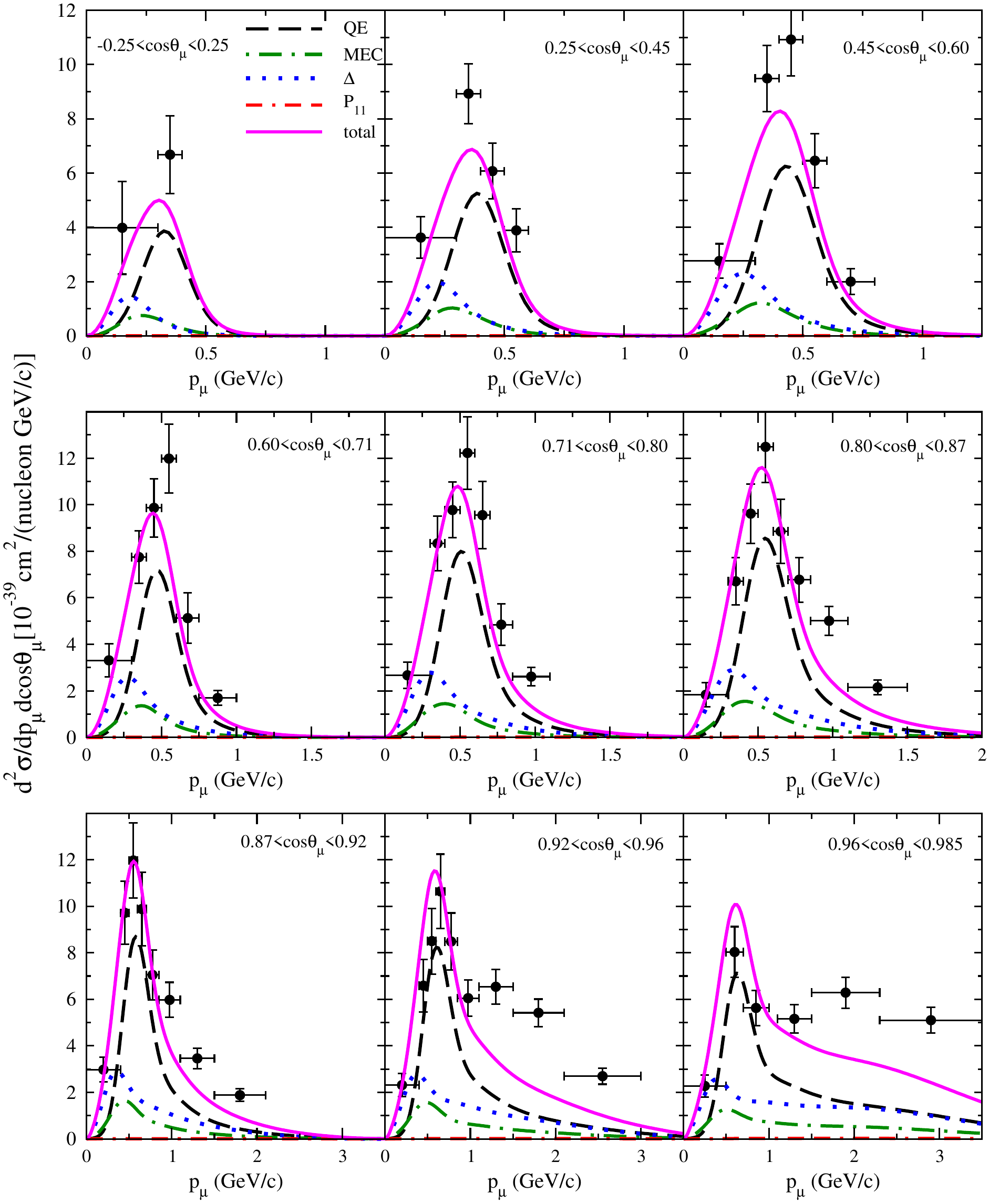}%
\caption{The SuSA  inclusive double differential $(\nu_\mu,\mu^-)$ cross section off $^{12}C$, averaged over the T2K flux, is displayed versus the muon momentum $p_\mu$. The separate QE, MEC, $\Delta$ and $P_{11}$ contributions  are shown.
Data  from Ref.~\cite{Abe:2013jth}. \label{fig:T2Kincl}}
\end{figure}

In Fig.~\ref{fig:T2Kincl} we compare the SuSA predictions with the T2K inclusive double differential $(\nu_\mu,\mu^-)$ cross section off $^{12}C$, displayed versus the muon momentum $p_\mu$ for different bins of the scattering angle $\theta_\mu$. The analysis of the separate QE, MEC, $\Delta$ and $P_{11}$ contributions, also shown in the figure, indicates that the $\Delta$ resonance gives a larger contribution than the MEC and is essential to explain the data, in particular at small $p_\mu$, whereas the contribution of the Roper resonance is totally negligible. Some disagreement with the data at large $p_\mu$ is observed for the most forward bin. This might be due to the lack of higher inelasticities in the model and will be explored in future work.

\begin{figure}[H]
\center
\includegraphics[width=12.5cm]{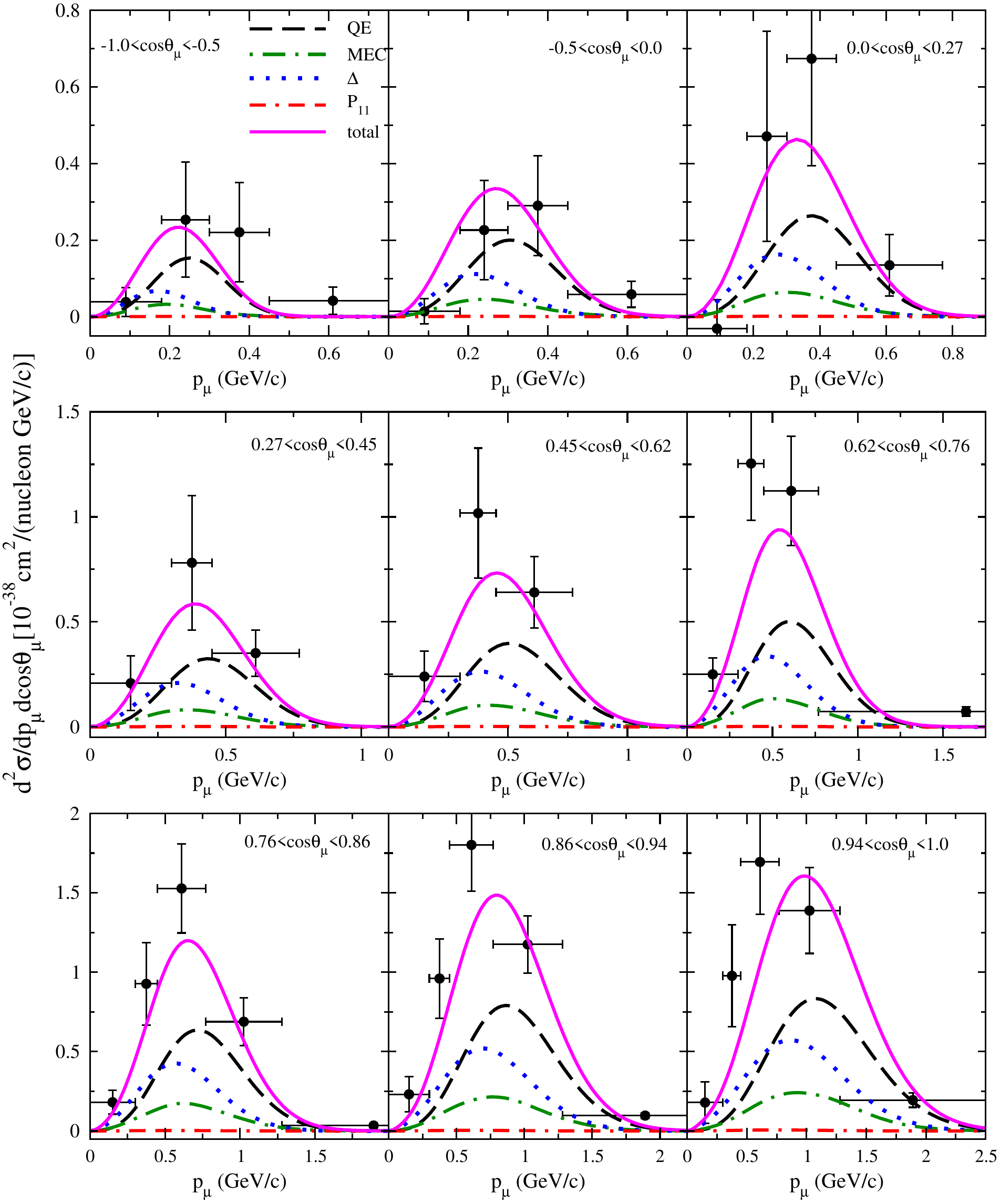}%
\caption{The SuSA  inclusive double differential $(\nu_\mu,\mu^-)$ cross section off $^{40}$Ar, averaged over the MicroBooNE flux, is  displayed versus the muon momentum $p_\mu$. The separate QE, MEC, $\Delta$ and $P_{11}$ contributions  are shown.
Data from Ref.~\cite{Abratenko:2019jqo}. \label{fig:MBincl}}
\end{figure}

Similar comments hold for the MicroBooNE inclusive cross section, shown in Fig.~\ref{fig:MBincl}. The comparison with these data is important to test the model for the argon nucleus, which will be the preferred target of future experiments. With respect to the T2K case (Fig.~\ref{fig:T2Kincl}) we observe a better agreement with the experimental result at high $p_\mu$ and an underestimation of the data at low $p_\mu$. The former is simply due to the larger errorbars in the experimental data, whereas the latter will likely be eliminated with the inclusion of higher inelasticities, that for MicroBooNE are expected to play a more important role due to the broader neutrino flux. Work along these lines is in progress.
As in the case of T2K, we stress that this is a preliminary work towards a more detailed and systematic comparison model/data. For this reason we chose not to calculate any $\chi^2$, but to just superimpose the theoretical curves to the experimental data in order to show qualitatively the successes and deficiencies of the model. A more quantitative and complete analysis will be performed in future work.

\begin{figure}[H]
\center
\includegraphics[width=12.5 cm]{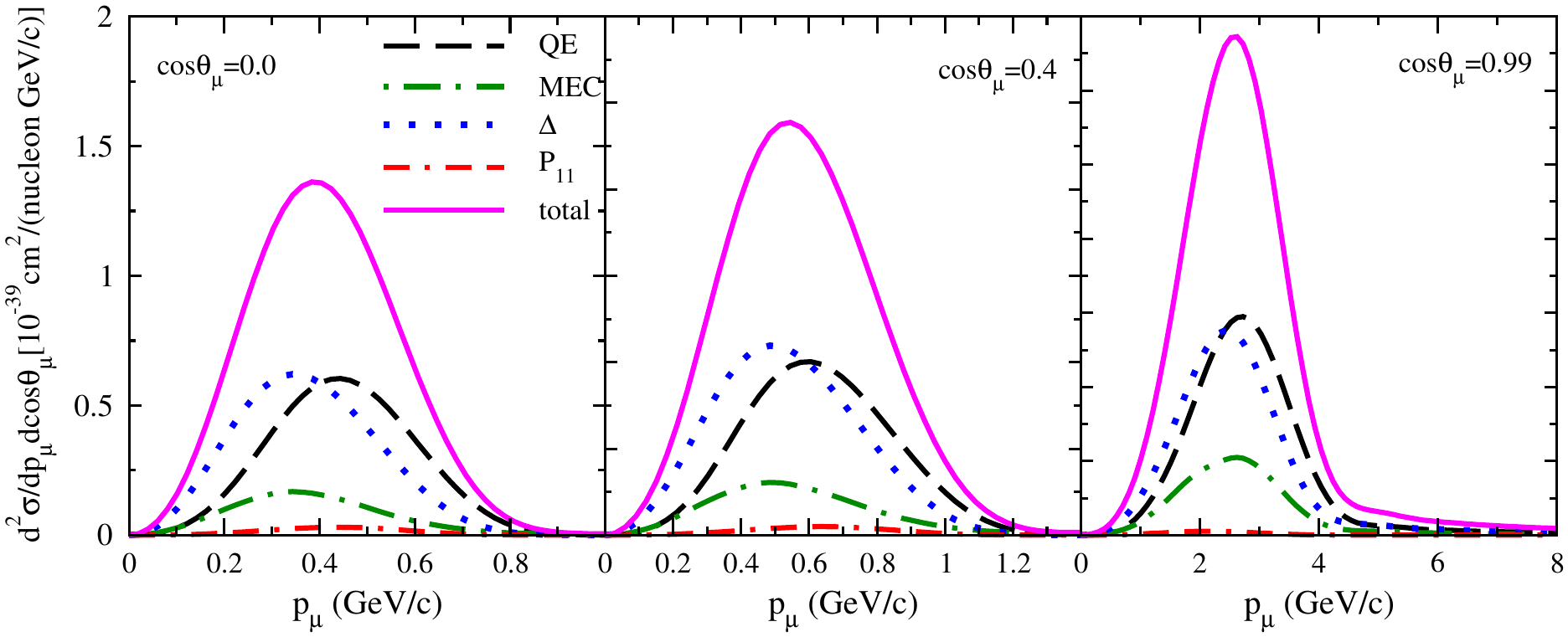}
\caption{The SuSA predictions for inclusive double differential $(\nu_\mu,\mu^-)$ cross section off $^{40}$Ar, averaged over the DuNE flux, are  displayed versus the muon momentum $p_\mu$. The separate QE, MEC, $\Delta$ and $P_{11}$ contributions  are shown. \label{fig:dune1}}
\end{figure}

\begin{figure}[H]
\center
\includegraphics[ width=8.5 cm]{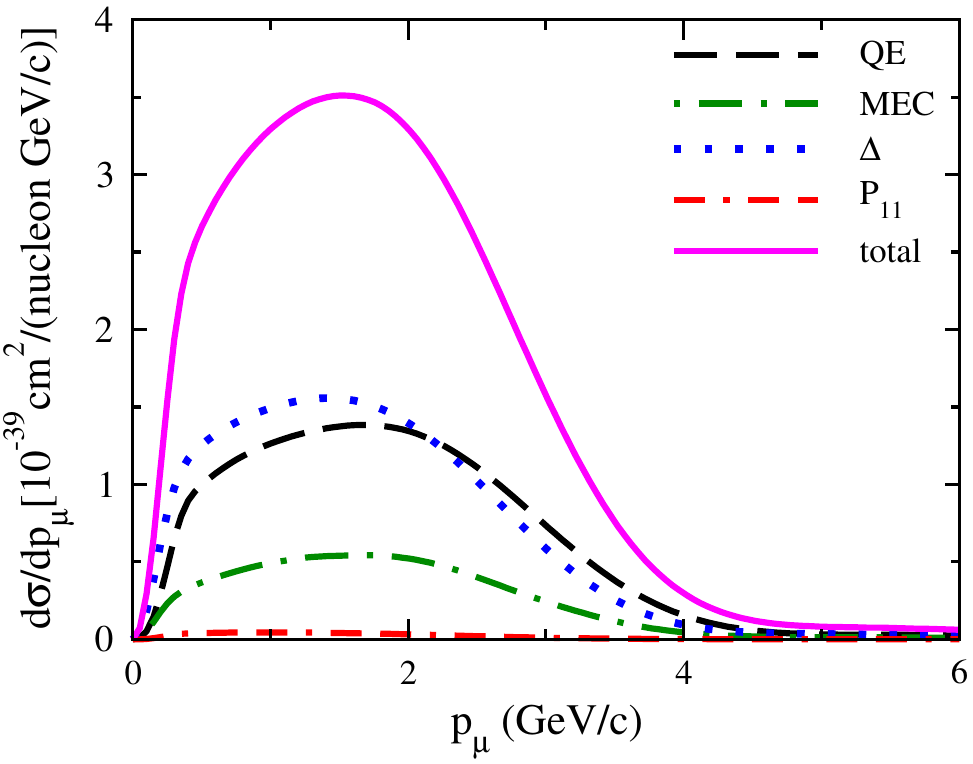}
\caption{The SuSA predictions for inclusive single differential $(\nu_\mu,\mu^-)$ cross section off $^{40}$Ar, averaged over the DuNE flux, are  displayed versus the muon momentum $p_\mu$. The separate QE, MEC, $\Delta$ and $P_{11}$ contributions  are shown. \label{fig:dune2}}
\end{figure}

Finally, in Figs.~\ref{fig:dune1} and \ref{fig:dune2} we present the predictions of the SuSA model for the future DUNE experiment,
 characterized by a higher energy and a broader  flux (see Fig.~\ref{fig:fluxes}). In this case the contribution of the $\Delta$ resonance becomes comparable to, or even larger than, the quasielastic one and the second resonance, $P_{11}$ plays a non-negligile, although small, role.

\section{Conclusions}
\label{sec:concl}

We have presented a unified treatment of the neutrino-nucleus response from the quasielastic up to the resonance region within a semi-phenomenological nuclear model (SuSA) based on the superscaling behaviour of inclusive electron scattering data. The approach is relativistic - as required by the kinematics - and, unlike the simpler relativistic Fermi gas model or other non relativistic models, it provides a good description of electron scattering data in a wide range of kinematics, a necessary test for models used in the analysis of neutrino oscillation experiments.
Moreover, the model is simple enough to be implementable in Monte Carlo generators used in the experimental analyses~\cite{Dolan:2019bxf}.

The SuSA model has been extensively studied in past work (see \cite{Amaro:2019zos} and references therein), with particular focus on the  quasielastic and 2p2h regions.
In this work for the first time the approach has been extended to study the first and second resonance regions, which will be of particular interest for the future high-energy experiment DUNE. 
The results of the model have been successfully compared with recent T2K and MicroBooNE data and predictions have been presented for DUNE.

Finally, it is worth pointing out that the contributions of heavier resonances  to the nuclear responses as well as interference effects should be taken into account in order to achieve a better quantitative description of the inelastic region. The present work represents a first step towards this more ambitious program.

\acknowledgments{This work was supported by the Istituto Nazionale di Fisica Nucleare under project
NucSys and by the University of Turin under Project BARM-RILO-20. The authors thank
G.D. Megias for providing the parameterization of the weak meson-exchange currents and
J.M. Franco Patino for useful comments on the results.}

\appendix
\section{Appendix}
\label{sec:app}
The $N\to P_{11}$(1440) form factors used in this work are
\begin{equation}
2F_1^* 
= \tau^* g_1^V
\,,\ \ 
2F_2^*=g_2^V
\,,\ \ 
G_A^*=g_1^A
\,,\ \ 
G_P^*=2g_3^A \,,
\end{equation}
where~\cite{Lalakulich:2006sw}
\begin{eqnarray}
g_1^V(q^2) &=& -\frac{4.6}{\left(1-\frac{q^2}{M_V^2}\right)^2\left(1-\frac{q^2}{4.3 M_V^2}\right)} \,,
\\
g_2^V(q^2) &=& +\frac{1.52}{\left(1-\frac{q^2}{M_V^2}\right)^2} \left[1-2.8 \ln\left(1-\frac{q^2}{1. {\rm GeV}^2}\right)\right] \,,
\\
g_1^A(q^2) &=& -\frac{0.51}{\left(1-\frac{q^2}{M_A^2}\right)^2\left(1-\frac{q^2}{3 M_A^2}\right)} \,,
\\
g_3^A(q^2) &=& \frac{m(m+m^*)}{m_\pi^2-q^2} g_1^A(q^2) \,,
\end{eqnarray}
with $M_V=0.84$ GeV and $M_A=1.05$ GeV.

\bibliographystyle{apsrev4-1}
\bibliography{bibliography}
\end{document}